# Pretransitional effects of the isotropic liquid-plastic crystal transition

A. Drozd-Rzoska[1], S. Starzonek[1], S. J. Rzoska[1*], [1]J. Łoś[1], Z. Kutnjak[2], S. Kralj[2,3]

[1]Institute of High Pressure Physics Polish Academy of Sciences, ul. Sokołowska 29/37,

01-142 Warsaw, Poland

[2] Condensed Matter Physics Department, Jožef Stefan Institute, Jamova 39,

1000 Ljubljana, Slovenia

[3] Faculty of Natural Sciences and Mathematics, University of Maribor, Koroška 160,

2000 Maribor, Slovenia

Corresponding authors

e-mail: sylwester.rzoska@unipress.waw.pl

**ABSTRACT**

We report on strong pretransitional effects across the isotropic liquid-plastic crystal melting temperature in linear and nonlinear dielectric response. Studies were carried out for cyclooctanol ($C_8H_{16}O$) in the unprecedented range of temperatures 120 K < $T$ < 345 K. Such pretransitional effects have not yet been reported in any plastic crystals. Results include the discovery of the experimental manifestation of the Mossotti Catastrophe behavior, so far considered only as a hypothetical paradox. The model interpretations of experimental findings are proposed. We parallel the observed pretransitional behavior with the one observed in octyloxycyanobiphenyl (8OCB), typical liquid crystal (LC), displaying a reversed sequence of phase transitions in orientational and translational degrees of order on varying temperature. Furthermore, in its nematic phase, we demonstrate first-ever observed temperature-driven crossover between regions dominated by isotropic liquid and smectic A pretransitional fluctuations. We propose a pioneering minimal model describing plastic crystal phase behavior where we mimic derivation of classical Landau-de Gennes-Ginzburg modelling of Isotropic - Nematic - Smectic A LC phase behavior.

## I. INTRODUCTION

Orientationally disordered crystals (ODICs)[1-3], also referred to as plastic crystals (PC), and liquid crystals (LCs)[4-5] are mesophases that can exist between the isotropic liquid and solid crystalline phases. Their unique properties are associated with the dominance of a single element of symmetry. For example, the nematic LC phase, representing the simplest LC configuration, is characterized by the orientational order and the translational disorder.[4-5] On the contrary, ODICs exhibit translational order and orientational disorder.[1-3] These configurations can be considered as convenient, simple and test-bed systems, from which fundamentals of the impact of different sequences of orientational and translational phase ordering on material properties could be extracted.[1-9] In particular, isotropic liquid (I) - mesophase (M) transitions are associated with melting/freezing of only a single element of symmetry.

However, existing literature suggests significantly different temperature-driven pretransitional behavior of I-ODIC and I-LC phase transitions. In the latter case, cooling towards the LC mesophase is associated with strong pretransitional effects in a wide temperature window within the liquid phase. Indeed, strong pretransitional changes of the Kerr effect (KE)[4-6, 10, 11] and Cotton Mouton effect (CME)[4-6, 12, 13] in the isotropic liquid phase of nematic LCs inspired development of the Landau-de Gennes (LdG) model[4-6, 14, 15], representing important corner-stone for theoretical description of LCs[3-7] and soft matter physics in general.[16, 17] On the contrary, in ODIC-forming systems the existing evidence indicates negligible weak pretransitional effect,[18-35] suggesting the behavior found in systems exhibiting classical liquid-crystal discontinuous phase transitions.[36-41] The only exception is reported for the optical Kerr effect (OKE) pretransitional effect of in *p*-terphenyl (1993),[42] which could be rather linked to the hidden isotropic-nematic (I-N) phase transition.

This report shows the first-ever experimental evidence of strong pretransitional effects for the isotropic liquid-ODIC (I-PC) phase transitions, extending over a wide temperature window. Such behavior is observed in dielectric constant and its strong electric field counterpart, the nonlinear dielectric effect (NDE), both surprisingly not tested so far. Phase behavior was probed in the extreme range of temperatures in cyclooctanol ($C_8OH$), one of the most classical ODIC-forming materials.[18-35] As a reference LC system, we studied a rod-like LC material, *n*-octyloxycyanobiphenyl (8OCB), which also revealed new aspects of pretransitional effects. Emerging similarities can offer a path for a common description of the melting phenomenon in LC- and ODIC-forming materials.

The paper is organized as follows. In *Sec. II* experimental methods used in our studies are introduced. In *Sec. III* we present experimental results in samples displaying liquid crystalline and plastic crystal phase ordering. Results are discussed in *Sec. IV*. In the last section we summarize results. Minimal models comparing liquid crystalline and plastic crystal phase behavior and derivation are given in *Appendix A*, whereas in *Appendix B* main the derivation of the classical Claussius-Mossoti equation is summarized.

## II. METHODS

In our study, we used the broadband dielectric spectrometer (BDS, Novocontrol), supported by the strong electric field facility enabling nonlinear dielectric spectroscopy studies and the Quattro temperature control unit.[43] Samples were placed in the flat-parallel measurement capacitor with plates made from Invar and gold-coated: diameter $2r = 20$ mm and the gap $d = 0.1$ mm. Scans of dielectric properties were carried out in the frequency range 0.1 Hz $< f <$ 10 MHz under the weak measuring voltage $U_{weak} = 1$ V, corresponding to the electric field $E_{weak} = 14$ kV/m. The scan of dielectric properties under the strong electric field was carried out for $U_{strong} > 1000$ V ($E_{strong} > 5$ MV/m), limited to $f < 10$ kHz.[38, 43]

The strong electric field related counterpart of dielectric constant is the nonlinear dielectric effect (NDE) [45-47]: $\varepsilon(E) = \varepsilon(E \to 0) + \Delta \varepsilon E^2 + ...$, where $\varepsilon(E \to 0) = \varepsilon$ represents the dielectric constant and for the nonlinear dielectric effect metric:

$$NDE := \frac{\Delta \varepsilon}{E^2} = \frac{\varepsilon(E) - \varepsilon}{E^2} \tag{1}$$

NDE was calculated using Eq. (1) from dielectric constant values in the middle of the static domain from $\varepsilon'(f, E)$ spectra as shown in Figure 1.

As the representative ODIC-type system we chose the cyclooctanol ($C_8H_{16}O$). Dielectric measurements were carried out in the extreme range of temperatures $120\ K < T < 345\ K$. Cyclooctanol is one of the most classical glass-forming materials, exhibiting the I-PC melting at $T_m \approx 278 - 292.5$ K, and subsequently the vitrification, for which the glass transition temperature $T_g$ is reported within the temperature window 150 K-240 K.[24-29] The material, with the highest declared purity, was purchased in Sigma-Aldrich. It was additionally dried using 4 Å molecular sieves.

Tests focused on the real part of dielectric permittivity and dielectric constant are hardly carried out for ODIC-forming materials so far. Examples of $\varepsilon'(f)$ spectra under the weak and strong electric fields are shown in Figure 1. The horizontal parts determine the dielectric constant: $\varepsilon = \varepsilon'(f)$. Under the weak electric field there is a notable impact of ionic dopants related to the notable increase for lower frequencies ($f < 100$ Hz), reflecting the Maxwell-Wagner process.[44] This phenomenon disappears under a strong electric field. The slowing down on cooling towards the glass transition causes that for the ODIC phase the impact of the high frequency relaxation process becomes detectable at lower and lower frequencies.

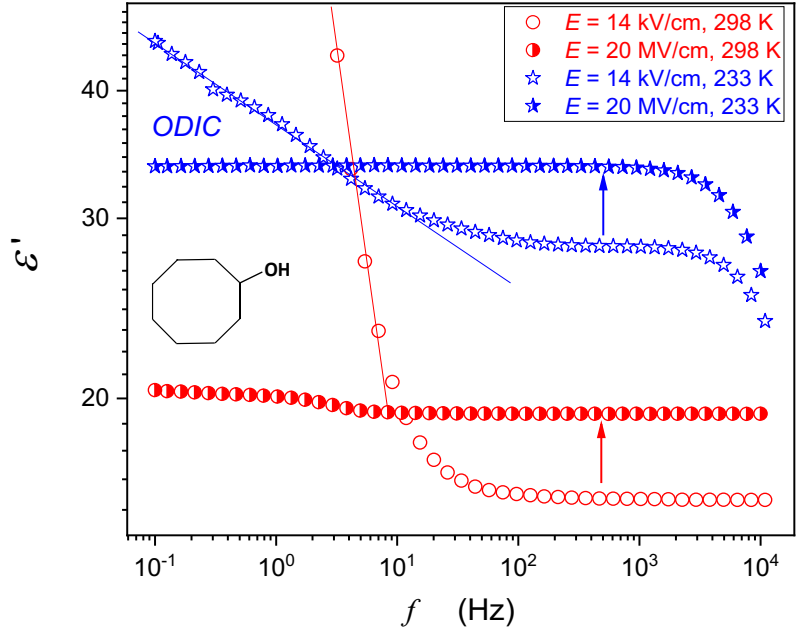

**Figure 1** The real part of dielectric permittivity ($\varepsilon'$) determined under the weak and the strong electric field in the (isotropic) liquid (red, circles) and the plastic crystal ODIC phase (blue, stars) in cyclooctanol. Horizontal parts in the plot yield dielectric constant $\varepsilon$ related static domain. The schematic sketch of cyclooctanol's structure is also shown.

As an LC representative system, we used octyloxycyanobiphenyl (8OCB). It consists of rod-like molecules with a relatively large permanent dipole moment $\mu \approx 6.6$ D parallel to the long molecular axis. It exhibits the following mesomorphism: isotropic liquid (I) - (352.7 K) - nematic (N) - (339.8 K) - Smectic A (SmA) - (322.1 K) - crystal (Cr). It belongs to the group of classical LC compounds, and it is included to LC mixtures for displays applications.[5, 6]

## III. RESULTS

We measured linear and nonlinear dielectric response in representative LC and ODIC materials, focusing on pretransitional phenomena. We first present the characteristic dielectric response of liquid crystalline 8OCB. Then a detailed study of plastic crystalline $C_8H_{16}O$ is subsequently presented.

## A. LC-forming materials: the case of 8OCB

When passing the isotropic liquid (I) - nematic (N) clearing (melting) temperature solely the orientational ordering freezes/melts whereas the translational fluid-like disarrangement remains. For phase transitions associated with smectic mesophases also some elements of a limited translational ordering appear.[5,6] Dielectric constant measurements reveal efficient macroscopic consequences of melting/freezing in LC materials. This physical property fingerprints effective dipole-dipole arrangements. Note that rod-like LC mesophases exhibit the head-to-tail invariance of the nematic director field $n$, which determines local uniaxial orientational LC order. If such LC compound molecule contains the permanent dipole moment parallel to the long molecular axis, then $\pm n$ invariance in the nematic phase favors cancellation of dipole moments.[5,6] The broadband dielectric spectroscopy is the basic tool also for determining dielectric constant as the stationary, frequency independent, domain of the reals part of dielectric permittivity ($\varepsilon'(f)$). The imaginary part of dielectric permittivity enables tests of dynamics associated with permanent dipole moments via the structural (primary, alpha) relaxation time determined from related loss curves peaks $\tau = 1/2\pi f_{peak}$.[48]

Figure 2 shows the temperature evolution of dielectric constant in the broad range of temperatures for 8OCB, encompassing isotropic, nematic and SmA phase regimes. Temperature changes of dielectric constant in the isotropic liquid phase are well portrayed by the relation:[49]

$$\varepsilon(T) = \varepsilon^* + a_\varepsilon(T - T_I^*) + A_\varepsilon(T - T_I^*)^{(1-\alpha)} \qquad (2)$$

describing dielectric response on lowering temperature. It holds $T > T_{I-N} = T_I^* + \Delta T_I^*$; $T_I^*$ is the temperature of the hypothetical continuous phase transition, $T_{I-N}$ determines the isotropic-nematic transition and $\Delta T_I^*$ is the metric of the discontinuity of the phase transition, $\varepsilon^*$, $a_\varepsilon$, $A_\varepsilon$ are phenomenological constants, and $\alpha \approx 0.5$ is the critical exponent. Note that the temperature window described well with Eq. (2) extends up to at least $T_{I-N} + 70K$.

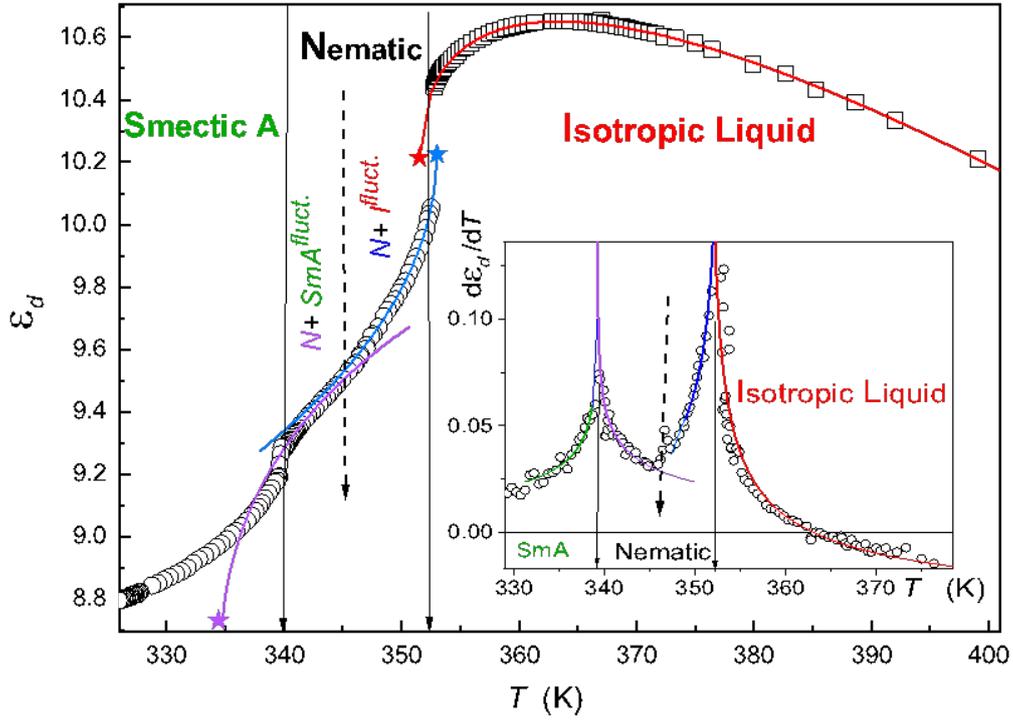

**Figure 2.** The temperature evolution of dielectric constant in the isotropic liquid and liquid crystalline mesophases of 8OCB. The latter is presented by the mean dielectric constant $\varepsilon_d = (2\varepsilon_\perp + \varepsilon_\parallel)/3$. Solid curves represent best fits using Eqs. (2-4). Stars denote the hypothetical extrapolated temperatures of continuous phase transitions. Solid arrows indicate temperatures of discontinuous I-N and N-SmA phase transitions. The dashed arrow indicates the crossover temperature in the nematic phase between domains dominated by isotropic liquid and SmA pretransitional fluctuations. The inset shows the distortions sensitive behavior of the derivative of mean dielectric constant, supporting the picture emerging from the main plot.

In the nematic phase, rod-like dipolar LC molecules are easily oriented an external field, then components of dielectric constant for molecules predominantly oriented perpendicularly ($\varepsilon_\perp$) and parellely ($\varepsilon_\parallel$) in the respect to the measuring electric field should be considered.[5,6] This was made by the strong magnetic field ($B \sim 1$ T). The dielectric anisotropy $\Delta\varepsilon = \varepsilon_\parallel - \varepsilon_\perp$, is commonly used to measure the order parameter in the nematic phase. In the nematic phase it holds that $\Delta\varepsilon(T) = \varepsilon^* + B(T_N^* - T)^\beta$: in 8OCB the classical tricritical value of the exponent $\beta \approx 1/4$ was reported.[59] Furthermore, the diameter of dielectric constant $\varepsilon_d = (2\varepsilon_\perp + \varepsilon_\parallel)/3$ is often of interest because it is related to the mean dielectric response of a dielectric constant in the nematic phase.[5,6] Its evolution in the nematic phase van be well portrayed by the relations:[50]

$$\varepsilon_d(T) = \varepsilon_N^* + a_N(T_N^* - T) + A_N(T_N^* - T)^{(1-\alpha)}, \qquad (3)$$

$$\varepsilon_d(T) = \varepsilon_{SmA}^* + a_{SmA}(T - T_{SmA}^*) + A_{SmA}(T - T_{SmA}^*)^{(1-\alpha)}. \qquad (4)$$

Here Eq. (3) and Eq. (4) are valid for temperature variations driving by $N \to I$ and $N \to SmA$ phase transitions, taking place at $T_{I-N}$ and $T_{N-SmA}$, temperatures. Quantities $\varepsilon_N^*, \varepsilon_{SmA}^*, a_N, A_N, a_{SmA}, A_{SmA}$ are phenomenological constants, $T_N^* = T_{I-N} + \Delta T_N$ and $T_{SmA}^* = T_{N-SmA} - \Delta T_{SmA}$ are temperatures of hypothetical continuous phase transitions determined from extrapolations from the nematic phase. Note that Eqs. (2-4) are linked to the same value of the critical exponent $\alpha \approx 1/2$.

We stress that Figure 2 reveals that the nematic phase is not homogeneous, the fact which has been (very) hardly indicated in the physics of liquid crystals so far.[5-16] There are two domains linked to vicinities of $N \to I$ and $SmA \leftarrow N$ phase transitions. This is particularly visible for the distortions sensitive plot shown in the inset in Figure 2, which also evidences the validity of ansatz used in fitting experimental measurements via Eqs. (3) and (4). Namely, it holds $\frac{d\varepsilon_d}{dT} \propto |T - T^*|^{-\alpha} \approx |T - T^*|^{-1/2}$, and $T^*$ stands either for $T_I^*$, $T_N^*$ or $T_{SmA}^*$, respectively. When explaining origins of pretransitional anomalies shown in Figure 2 one should indicated that in the isotropic liquid phase dielectric constant of prenematic fluctuations is much smaller than for the isotropic liquid surrounding $\varepsilon_{fluct.} \ll \varepsilon_{surr.}$. Core increase of the volume occupied by prenematic fluctuation occurring for $T \to T_{I-N}$ leads to the crossover $d\varepsilon(T)/dT < 0 \to d\varepsilon(T)/dT > 0$. For $N \to I$ transition the isotropic fluctuations are surrounded by the nematic background, thus $\varepsilon_{fluct.} \gg \varepsilon_{surr.}$, leading to Eq. (3). For $SmA \leftarrow N$ the distortion of the orientational ordering by 1-D smectic ordering causes that $\varepsilon_{fluct.} > \varepsilon_{surr.}$, yielding Eq. (4).

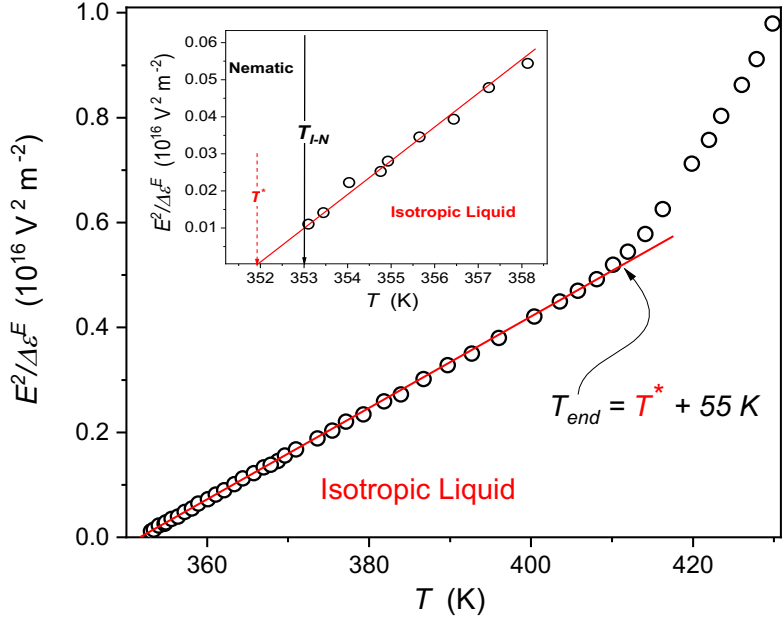

**Figure 3.** The temperature dependence of the reciprocal of NDE, strong electric field related changes of dielectric constant, in the isotropic liquid phase of liquid crystalline 8OCB. The line shows the validity of portrayal via Eq. (5), extending from the Isotropic-Nematic melting temperature to $T_{end}$. The inset shows the behavior in the immediate vicinity of the I-N melting (clearing) temperature, indicated by the solid arrow. The dashed arrow indicates the hypothetical continuous phase transition.

The temperature evolution of the strong electric field ($E$) related (nonlinear) counterpart of dielectric constant, is shown in Figure 3. This magnitude, known as the nonlinear dielectric effect (NDE), exhibits an extreme pretransitional anomaly in the isotropic liquid phase of LC materials, portrayed by the following relation:[51-55]

$$\frac{\Delta\varepsilon^E}{E^2} = \frac{C}{T - T_I^*} \tag{5}$$

where $C$ is the material constant and $T_I^*$ is the temperature of the hypothetical continuous phase transition as discussed above.

Parallel relations describe the behavior of the intensity of the scattered light, Cotton-Mouton effect or the Kerr effect in the isotropic liquid phase of nematic LC materials.[5-16] In fact, those studies were the inspiration for the Landau-de Gennes model (LdG), the key theoretical concepts for the physics of liquid crystals,[5,6] and for the general soft matter physics.[16,17] Note that NDE is the only method for which Eq. (5) also obeys for I-N, I-SmA, I-SmE, I-N* transitions. This is associated with the fact that for NDE the measurement time scale, related to the radiofrequency of the weak measuring field $\tau_{method} \gg \tau_{fl.}$.[51-55] For KE, CME or IL $\tau_{method} \ll \tau_{fl.}$, what is associated with the light – related measurement frequency. Eq. (5) can be derived from LdG model-based consideration and also from the expression originally derived to model NDE and KE pre-critical effects on approaching the critical consolute point in binary mixtures of limited miscibility or the gas-liquid critical point:[56-58]

$$\frac{\Delta \varepsilon^E}{E^2} \propto \chi \langle \Delta M^2 \rangle_V \qquad (6)$$

where $\langle \Delta M^2 \rangle_V \propto |T-T_C|^{2\beta}$ stands for the averaged square of the order parameter fluctuations and $\chi = \chi_0 |T-T_C|^{-\gamma}$ denotes the compressibility (order parameter related susceptibility). The critical temperature $T_C$ is associated with the critical temperature of a continuous phase transition. For the isotropic phase of LC materials, it then holds $T_C = T_I^*$.

In the isotropic liquid phase, the mean-field approximation works relatively well, owing to the elongated and rod-like form of LC molecules which increases the number of neighboring molecules. It is notable that both compressing and strong electric field can change the volume/shape of pretransitional fluctuations but cannot influence dielectric constant related to prenematic fluctuations. Consequently, in the isotropic liquid phase, it is reasonable to set $\langle \Delta M^2 \rangle_V \propto (\Delta \varepsilon)^2 = const$ and $\chi = \chi_0/(T-T^*)^{-\gamma}$ with the classical exponent $\gamma = 1$, what leads to Eq. (5). Notable, that in the isotropic phase changes in dielectric constant are well described using Eq. (2) up to at least 70 K above the clearing temperature. For NDE the pretransitional effect described by Eq. (5) persist till $T_{end} \approx T_I^* + 40K$.[51-55] This can be linked to the reduction of pre-mesomorphic fluctuations to 2-3 molecules, and what make their detection by methods directly coupled to their presence, such as NDE, impossible.

### B. ODIC-forming materials: the case of cyclooctanol

For plastic crystalline materials when cooling below the I-ODIC freezing/melting temperature $T_m$ the translational ordering appears but the orientational freedom remains. Experimental evidences from previously published results[18-35] seems to be clear: there are no pretransitional behavior in the surrounding of $T_m$, contrary to the discussed above case of LC materials. Note that on cooling ODIC forming materials most often terminate in the orientationally disordered solid glass state. Consequently, studies of such systems are focused mainly on the glass transition problem, for which the enormous shift of the primary relaxation time from pico-/nanoseconds to $\tau(T_g) \approx 100$ s at the glass temperature $T_g$ is the key artifact. Consequently, the broad band dielectric spectroscopy (BDS) is an appropriate experimental research tool to study main features of ODICs.

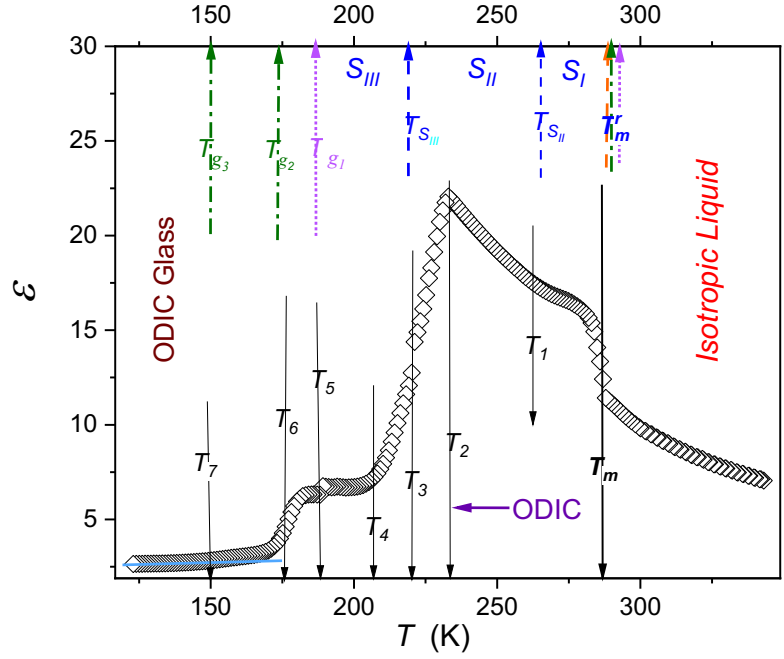

**Figure 4.** Temperature evolution of the dielectric constant in ODIC-forming cyclooctanol. Solid arrows pointing downwards show characteristic temperatures extracted from our measurements: $T_m = 286.4$ K, $T_1 = 233.$ K, $T_2 = 207.1$ K, $T_3 = 178.2$ K. Arrows pointing upwards are related to previously published values of relevant critical temperatures: (i) blue, dashed,[25] (ii) dotted, pink,[27] and (iii) dot-dashed, green.[28]

Figure 4 shows the temperature evolution of dielectric constant of cyclooctanol ($C_8OH$), one of the most classical ODIC-forming materials, in the broadest ever studied temperature range covering also the liquid phase from ca. 120 K up to $T_m + 70$ K. Such extreme range has to be associated with the qualitative shift of the structural (primary) relaxation time from $\tau \approx 5$ ps to $\tau(T_g) \approx 100$ s, what also leads to the shift of the location of the static domain of $\varepsilon'(f) = \varepsilon$ from ~ 1 MHz to even ~1 Hz domain. This shift was considered for the experimental data discussed below. Figure 4 shows the temperature evolution of dielectric constant for cyclooctanol. Previous investigations focused mainly on heat capacity and BDS based dielectric relaxation studies and led to the identification of three possible ODIC phases ($S_I$, $S_{II}$, $S_{III}$) and suggested three possible glass temperatures ($T_{g1}$, $T_{g2}$, $T_{g3}$). These values are shown in Figure 4 by dashed arrows, oriented upwards. The scan of the dielectric constant sheds new light on ODIC mesomorphism and characteristic temperatures in C$_8$OH. There are also characteristic temperatures not specified so far, such as $T_2$ associated with the sharp crossover $d\varepsilon/dT > 0 \leftarrow d\varepsilon/dT < 0$ and $T_4$ linked to the change $d\varepsilon/dT \approx 0 \leftarrow d\varepsilon/dT > 0$. The question arises for $T_{S_{II}}$, indicated earlier as the phase transition between $S_I$ and $S_{II}$ ODIC phase. As shown below, $T_{S_{II}} \approx T_1$ can be associated with crossover between two pretransitional domains in the ODIC phase. Worth recalling is a similar crossover occurring in the nematic phase of liquid crystalline 8OCB, see Figure 2. Changes of $\varepsilon(T)$ values can be associated with the increase/decrease of the freedom of permanent dipole moment for following changes in the electric field. The behavior described by $d\varepsilon/dT < 0$ or $d\varepsilon/dT > 0$ indicates the preference for parallel or antiparallel arrangements of permanent dipole moments.[45]

Figure 5 shows the temperature evolution of the reciprocal of dielectric constant, which reveals simple forms of the temperature evolutions in the isotropic liquid for $T > T_m$ and in the ODIC phase for $T_2 < T < T_m$:

$$\varepsilon^{-1}(T) = b + aT \quad \rightarrow \quad \varepsilon(T) \propto \frac{A}{T - T^+}, \tag{7}$$

where $a, b, A$ are constants and $T^+$ denotes the extrapolated singular temperature; the linear regression analysis yields in the liquid phase $T^+ = 195$ K ($A = 1078$ K), and $T^+ = 115$ K ($A = 2490$ K) in the ODIC phase.

Such a description is also validated by the distortions-the sensitive derivative plot $d\varepsilon^{-1}(T)/dT$ shown in the upper inset in Figure 5. Its parameterization shows that in the immediate vicinity of $T_m$ the additional critical-like behavior appears:

$$\frac{d\varepsilon}{dT} \propto (T - T^*)^{-1.5} \tag{7a}$$

for $T = T_m \pm 10K$ and singular temperatures $T^* = 276.0$ K for $I \rightarrow ODIC$ and $T^*=293.8$ for $ODIC \rightarrow I$ transitions.

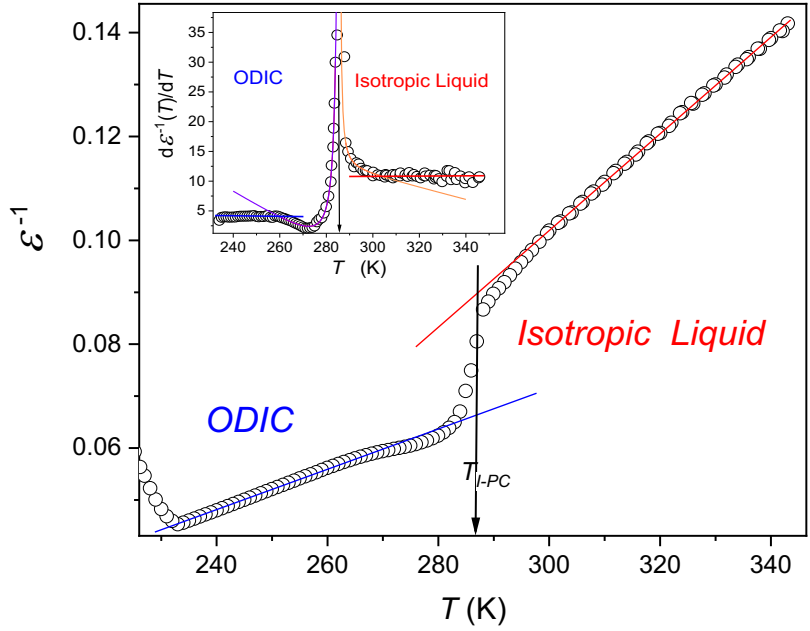

**Figure 5.** The evolution of the reciprocal of dielectric constant in cyclooctanol. The plot is based on experimental shown in Figure 4. In the isotropic liquid and in the notable part of the ODIC phase, the $\varepsilon^{-1}$ temperature dependence exhibits linear temperature dependence, suggesting the validity of Eq. (7) with extrapolated singular temperatures $T^+ = 195$ K (blue line) and $T^+ = 115$ K (red line). The derivative-based and distortions sensitive analysis in the upper inset confirms the mentioned behavior, also revealing a distortion in the very immediate vicinity of $T_m$. It can be portrayed by Eq. (7a), what is shown by solid curves.

Nonlinear dielectric effect (NDE) describes changes of dielectric constant under the strong electric field. It can directly detect the appearance of collective phenomena, such as pretransitional fluctuations. This sensitivity is associated with different interactions of the collective species (fluctuation) and the surrounding background with the strong electric field. Figure 6 presents results of the first-ever NDE measurements in the liquid and ODIC phases of aplastic crystal-forming material, $C_8OH$ in the given case.

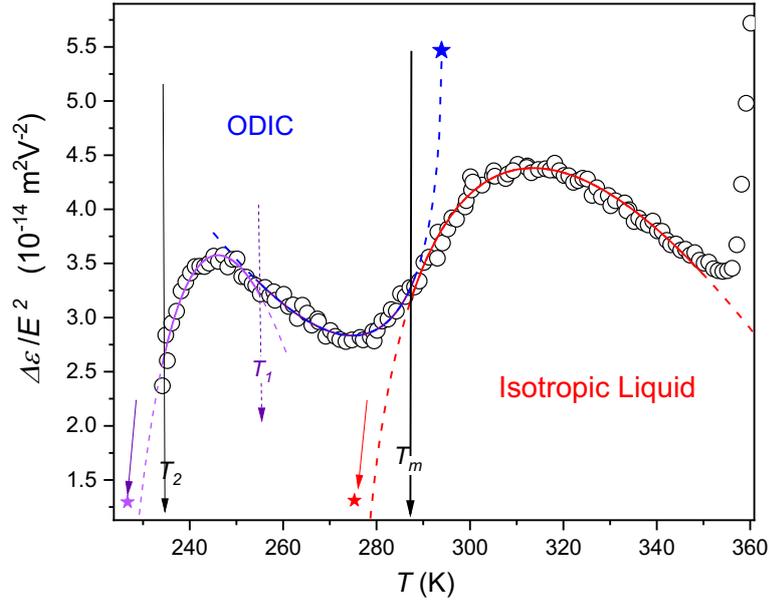

**Figure 6.** The temperature dependence of nonlinear dielectric effect (NDE) in the liquid and ODIC phases of cyclooctanol. Solid curves are related to Eq. (8) with $T_m$ = 287.5 K ($T^*$ = 276 K), $T_1$ = 254.5 K ($T^*$=293.8), $T_2$ = 234 K ($T^*$=220 K).

**TABLE I**

Values of parameters describing *NDE* vs. *T* experimental data in the liquid and ODIC phases of cyclooctanol (Figure 6) when portraying by Eq. (8).

| Transition → Parameters ↓ | ODIC → postODIC | ODIC → I | I → ODIC |
|---|---|---|---|
| $\varepsilon^*$ | -10.930 | 5.467 | -1.630 |
| $A$ | -0.556 | 0.138 | -0.160 |
| $B$ | 5.682 | -1.205 | 1.963 |
| $T^*$ (K) | 220.0 | 293.9 | 276.0 |
| $\phi$ | 0.5 | 0.5 | 0.5 |
| $T_m$ (K) | 232.4 | 287.2 | 287.2 |
| $\Delta T^*$ (K) | 12.4 | 6.7 | 11.2 |

There is a strong and long-temperature range pretransitional effect in the liquid phase, extended even up to ca. $T_x \approx T_m + 50$ K, well portrayed by the relation:

$$\frac{\Delta \varepsilon}{E^2}(T) = \frac{\Delta \varepsilon}{E^2}(T^*) + A_\Delta |T - T^*| + B_\Delta |T - T^*|^\phi \tag{8}$$

where the exponent $\phi = 1/2$ and $T^*$ is the singular temperature, possibly that associated with a hypothetical hidden continuous phase transition. The value of $\Delta T^* \approx 12 K$ can be considered as the metric of $ODIC \leftarrow I$ phase transition discontinuity. Notable, that this value is similar to ones detected for I-SmE transitions in liquid crystalline material via NDE measurements.[47,55] It is worth recalling that the SmE phase the orientational arrangement is assisted by so complex translational structure that such materials are encountered both to liquid crystalline and generalized plastic crystalline materials.[6,55]

Note, that for $ODIC \leftarrow I$ transition approximately the same value of the singular temperature $T^* \approx 276$ K were obtained from NDE studies (Eq. (8)) and from subtle changes of dielectric constant in the immediate vicinity of $T_m$ (Eq. (7a)). The comparison of discontinuities of phase transitions yields $\Delta T^* (ODIC \rightarrow I) \approx (1/2) \Delta T^* (I \rightarrow ODIC)$. This value is similar to the ones noted for the $I \rightarrow N$ and $N \rightarrow I$ phase transitions in liquid crystalline materials.[50]

## IV. DISCUSSION

Figures 5 and 6 reveal a strong pretransitional anomalies in liquid and plastic crystal phases of cyclooctanol. We claim it can announce a phase transition in orientational order of cyclooctanol molecules. This ordering is also linked to the collective ordering of permanent dipoles that the molecules host. Below we illustrate that this phase transition results in the Mossotti catastrophe.[20, 59, 60, see also *Appendix*] The latter is fingerprinted in the observed critical pretransitional behavior in dielectric responses.

We originate from the classical Claussius-Mossotti equation, which relates the macroscopic dielectric response with microscopic dipolar properties (see *Appendix B*). We set that the system consists of two different types of electric dipoles, to which we refer as collective and non-collective dipoles, respectively. The former (latter) behave as effectively coupled (non-coupled) ensemble. The total electric polarization $\boldsymbol{P}$ of the system is then expressed as

$$\boldsymbol{P} \sim n^{(c)}\boldsymbol{p}^{(c)} + n^{(n)}\boldsymbol{p}^{(n)} = n^{(c)}\alpha^{(c)}\boldsymbol{E}_{loc}^{(c)} + n^{(n)}\alpha^{(n)}\boldsymbol{E}_{loc}^{(n)}, \tag{9}$$

where the superscripts $^{(c)}$ and $^{(n)}$ refer to collective and non-collective contributions, $\alpha$ is the polarizability constant, $n$ labels volume density of electric dipoles, $\boldsymbol{p}$ stands for an average electric dipole experiencing the local electric field $\boldsymbol{E}_{loc}$. Furthermore, we assume $\boldsymbol{E}_{loc}^{(c)} \sim \boldsymbol{E}_{loc}^{(n)} \equiv \boldsymbol{E}_{loc} = E_{loc}\boldsymbol{e}_E$, where $|\boldsymbol{e}_E| = 1$.

We set that the collective dipolar response exhibits temperature driven phase transition at the critical temperature $T_c$. We define the orientational order parameter amplitude $m_o$ of the phase transition by

$$m_o = <\boldsymbol{e} \cdot \boldsymbol{e}_E> = <cos\theta>. \tag{10}$$

where $\boldsymbol{e}$ denotes a temporal orientation of a dipole, <….> stands for the ensemble average, and $\theta$ is the angle between the unit vectors $\boldsymbol{e}$ and $\boldsymbol{e}_E$. Therefore, isotropic distribution of $\boldsymbol{e}$ or their strict alignment along the symmetry breaking direction $\boldsymbol{e}_E$ results in $m_o$=0 or $m_o$=1, respectively. Consequently, an average collective electric dipole can be expressed as

$$p^{(c)} = p_0^{(c)} m_o. \tag{11}$$

The corresponding free energy density $f$ describing the critical behavior is given by $f = f_c + f_E$. It consists of the condensation ($f_c$) and the external field ($f_E$) contributions:

$$f_c \sim a_0(T - T^*)m_0^2 - bm_0^4 + cm_0^6 \tag{12a}$$

$$f_E \sim -n^{(c)} E_{loc} p_0^{(c)} m_o. \tag{12b}$$

Here we use the simplest possible modeling to illustrate the mechanism yielding the observed pretransitional behavior. The more detailed description is presented in *Appendix A*. The condensation term describes the temperature-driven critical behavior of collective dipoles for $E_{loc} = 0$. The quantities $a_0 > 0$ and $b$ are the Landau expansion coefficients. For $b>0$ the phase transition is discontinuous and $T^*$ determines the supercooling temperature.

A finite value of $E$ (and consequently $E_{loc}$) enforces a finite macroscopic dipolar orientational ordering in the whole temperature regime. For relatively high temperatures it holds

$$f \sim a_0(T - T^+)m_0^2 - n^{(c)}p_0^{(c)}E_{loc}m_o. \tag{13}$$

The minimization of the above relation with respect to $m_o$ yields

$$m_o = \frac{n^{(c)}p_0^{(c)}E_{loc}}{2a_0(T-T^+)}. \tag{14}$$

Considering Eq. (9) it follows $\alpha^{(c)} = \frac{n^{(c)}p_0^{(c)2}}{2a_0(T-T^+)}$. With this in mind we generalize the classical Claussius-Mossotti equation (see Eq. (B4) in Appendix B) to the case consisting of collective and non-collective dipolar contributions. By considering $n\alpha \to n^{(c)}\alpha^{(c)} + n^{(n)}\alpha^{(n)}$, it follows

$$\frac{\varepsilon-1}{\varepsilon+2} = \frac{n^{(n)}\alpha^{(n)}}{3\varepsilon_0} + \frac{n^{(c)2}p_0^{(c)2}}{6\varepsilon_0 a_0(T-T^+)}, \tag{15}$$

which predicts a critical-like response for temperatures $T > T^+$. The equation exhibits the Claussius-Mossotti catastrophe at $T = T^+$.

In the above description we considered only the phase transition in the orientational degree of order. However, in the studied ODIC system there is a sequence of phase transitions in translational degrees of freedom in the temperature regime above $T_c$. In general the translational and orientational degrees are weakly coupled (see *Appendix A*), which manifests in different measured values of $T^+$ above and below $T_m$, as evidences Figure 5. Furthermore, the above description assumes ferroelectrically ordered orientationally ordered phase.

However, the ordered phase could also exhibit an anti-ferroelectric order. In this case $m_o$ refers to ordering in an anti-ferroelectric subdomain.

Furthermore, ODIC and 8OCB LC display reversed temperature dependent behavior of linear and nonlinear dielectric responses in the temperature regime $T > T_c$ ($T_c \equiv T_{I-N}$ for 8OCB and $T_c \equiv T_m$ for $C_8H_{16}O$). In 8OCB the low field (linear) dielectric response exhibits non-monotonic $\varepsilon(T)$ dependence, switching from $\frac{d\varepsilon}{dT} < 0$ to $\frac{d\varepsilon}{dT} > 0$ behavior. On the contrary, the high field NDE (nonlinear) response monotonically increases. In ODIC one observes opposite behavior. Namely, $\varepsilon(T)$ monotonically increases, while NDE exhibits non-monotonic temperature behavior on decreasing $T$.

A possible qualitative explanation for the observed pretransitional behaviors in 8OCB is as follows. In both linear and nonlinear regime, the system consists of a "sea" of strongly fluctuating (but weakly ordered by $E$) LC molecules within which exist "islands" (clusters) exhibiting relatively strong paranematic order. The former molecules effectively behave like non-collective system members. Their orientational order is dominated by thermal fluctuations. On the contrary, paranematic clusters correspond to relatively ordered domain regions, each characterized by an average local symmetry breaking direction $\boldsymbol{n}^{(c)}$. These regions contribute to the collective response, which is dominated by nematic interactions. These favor locally parallel mutual orientation (exhibiting head-to-tail invariance) of neighboring LC molecules. We henceforth label quantities referring to non-collective and collective contributions with superscripts (n) and (c), respectively. Hence, we express the total dielectric response as

$$\varepsilon = \frac{V^{(n)}}{V}\varepsilon^{(n)} + \frac{V^{(c)}}{V}\varepsilon^{(c)}, \tag{16}$$

where $V = V^{(n)} + V^{(c)}$ determines the total volume of a sample. On decreasing $T$ the relative presence of collective contribution increases. Namely, it roughly holds $V^{(c)} \sim N\xi_n^3$, where $N$ determines the number of nucleated paranematic clusters, and $\xi_n$ is the nematic order parameter correlation length which increases on approaching $T_c$.

In the linear regime, the orientational probability distribution function of $\mathbf{n}^{(c)}$ is roughly isotropic. Consequently, we claim that for temperatures $T > T_c$ it holds $\varepsilon^{(c)} < \varepsilon^{(n)}$. Namely, $\mathbf{E}$-driven reorientations of electric dipoles of non-collective molecules are on average not hindered by LC molecular fields. On the other hand, a relatively larger number of paranematic clusters is oriented perpendicular to the weakly imposed symmetry breaking field $\mathbf{E}$. The response of such molecules is closer to $\varepsilon_\perp$, where $\varepsilon_\perp \ll \varepsilon_\parallel$, and $\varepsilon_\perp$ ($\varepsilon_\parallel$) measures the response for LC molecules for fields applied perpendicular (along) the nematic director field. Note that values of $\varepsilon_\perp$ and $\varepsilon_\parallel$ for temperatures $T > T_c$ are smaller (due to higher temperatures and finite sizes of paranematic clusters) but comparable to those in the nematic phase. With this in mind it follows that on decreasing $T$ the relative contribution of $\varepsilon^{(c)}$ progressively dominates in $\varepsilon$ and consequently $\varepsilon(T)$-slope gradually converts from $\frac{d\varepsilon}{dT} < 0$ to $\frac{d\varepsilon}{dT} > 0$.

On the contrary, in the nonlinear regime, the external field is strong enough to align $\mathbf{n}^{(c)}$ of most of clusters along $\mathbf{E}$. Consequently, in this regime the $\varepsilon^{(c)}$ response is dominated by $\varepsilon_\parallel$ contribution, where $\varepsilon_\parallel \gg \varepsilon_\perp$. This yields monotonically increasing dielectric response on decreasing $T$. In the case of ODIC situation is much more complicated. Namely, both pretransitional clusters in translational and orientational order are expected.

## V. CONCLUSIONS

The advanced way of dielectric constant determination from BDS spectra matched with the extreme range of temperatures studied (120 K < $T$ < 350 K, for 220 selected temperatures) and supported nonlinear dielectric effect studies led to the qualitatively new insight into the physical properties of ODIC-forming cyclooctanol. Such unique experimental results led to the evidence for a well-defined pretransitional effect both for the isotropic liquid-ODIC transition and within the ODIC phase. To confirm these results additional NDE measurements have been carried out using the dual field measurement principle, being the extreme resolution method associated with strong dielectric field pulses limited to only a few milliseconds.

The analysis of the reciprocal of dielectric constant vs. temperature showed that ODIC-forming materials, both in the liquid and ODIC phase, can constitute the system were the Mossotti catastrophe[50, 56] is the experimental fact, not only the interesting speculative paradox. Consequently, the question arises of designed studies on ODIC-forming systems can lead to a new type of ferroelectric material?

Finally, this report indicates the possible significance of studies in symmetry-selected systems for approaching the cognitive breakthrough for the puzzling case of discontinuous phase transitions.

**APPENDIX A: ODIC MINIMAL MODEL**

Below we propose a minimal model for temperature-driven liquid-plastic crystal-crystal phase transition sequence, which we henceforth refer as the ODIC model. We introduce it based on the analogous behavior in ordinary thermotropic liquid crystals, while in the ODIC case we reverse the phase transition sequence in which an orientational and translational order appear. We first recall a minimal model describing the thermotropic isotropic (I) - nematic (N) - smectic A (SmA) liquid crystal phase sequence on decreasing temperature. Then we present our minimal ODIC model.

**Liquid crystals**

We consider bulk thermotropic LCs, which exhibit on lowering temperature liquid (isotropic), orientational nematic (N) order, and smectic A (SmA) order. For this purpose, we use a simple Landau-de Gennes-type[4-6] uniaxial mesoscopic description. The uniaxial orientational nematic order is described by the uniaxial tensor nematic order parameter

$$\boldsymbol{Q} = S(\boldsymbol{n}\otimes\boldsymbol{n} - \boldsymbol{I}/3), \qquad (A1)$$

consisting of the uniaxial order parameter $S$ and the nematic director field $\boldsymbol{n}$. Here $S \in [-1/2, 1]$ reveals the degree of nematic ordering. Note that states $S = \pm 1$ are physically different, and $S=0$ fingerprints the isotropic order. The unit vector field $\boldsymbol{n}$ exhibits the head-to-tail invariance, i.e., the states $\pm \boldsymbol{n}$ are physically equivalent. The smectic A translational order is determined by the complex smectic order parameter

$$\psi = \eta e^{i\phi}. \qquad (A2)$$

The amplitude $\eta \geq 0$ measures the degree of translational order. The position of smectic A layers is determined by the phase $\phi$. In the bulk nematic equilibrium, $S(\boldsymbol{r})$ and $\boldsymbol{n}(\boldsymbol{r})$ are spatially homogeneous, where $\boldsymbol{n}$ is pointing along an arbitrary symmetry-breaking direction. In the bulk equilibrium SmA phase, in addition to the spatially homogeneous nematic order, also a regular stack of smectic layers appears. It is characterized by a spatially homogenous value of $\eta(\boldsymbol{r})$ and the phase $\phi(\boldsymbol{r}) = q_0 \boldsymbol{n}.\boldsymbol{r}$ determines position of layers of thickness $d_0 = \frac{2\pi}{q_0}$. Note that $\boldsymbol{n}$ points along a smectic layer normal, i.e. $\boldsymbol{n} = \frac{\nabla\phi}{|\nabla\phi|}$.

At the I-N and the N-SmA phase transition, a continuous symmetry is broken in orientational and translational order, respectively. Consequently, both order parameters fields $(\boldsymbol{Q}(\boldsymbol{r}), \psi(\boldsymbol{r}))$ consist of two qualitatively different contributions: the *amplitude* fields $(S(\boldsymbol{r}), \eta(\boldsymbol{r}))$ and the *symmetry breaking* (also referred to as the *gauge*) fields $(\boldsymbol{n}(\boldsymbol{r}), \phi(\boldsymbol{r}))$. The *amplitudes* reveal strengths of the established ordering, while the symmetry-breaking fields reveal symmetry breaking choice of a relevant phase transition. Consequently, in a bulk equilibrium phase, a relevant *amplitude* field exhibits unique value, while a relevant *gauge field* exhibits infinite degeneracy.

In terms of the nematic and smectic order parameters, we express the free energy $f = f_n^{(c)} + f_n^{(e)} + f_n^{(f)} + f_s^{(c)} + f_s^{(e)} + f_c$ as a sum of nematic and smectic contributions.

These terms are (in the lowest order expansion necessary to describe a sequence of I-N and N-SmA phase transitions) commonly expressed as follows:[16]

$$f_n^{(c)} = a_n(T - T_n^*)S^2 - b_n S^3 + c_n S^4, \tag{A3a}$$

$$f_n^{(e)} = L|\nabla \mathbf{Q}|^2, \tag{A3b}$$

$$f_n^{(f)} = -\varepsilon_0 \Delta\varepsilon S(\mathbf{n}.\mathbf{E})^2/2, \tag{A3c}$$

$$f_s^{(c)} = a_s(T - T_s^*)|\psi|^2 + b_s|\psi|^4 + \cdots, \tag{A3d}$$

$$f_s^{(e)} = C_\perp |(\mathbf{n} \times \nabla)\psi|^2 + C_\parallel |(inq_0 - \nabla)\psi|^2. \tag{A3e}$$

$$f_c = -DS\eta^2. \tag{A3f}$$

The quantities $a_n, b_n, c_n, T_n^*, a_s, b_s, T_s^*, D$ are positive material constants enabling a 1st order I-N and N-SmA phase transition. The character of the latter transition depends on $D$ value which, determines the coupling strength between $S$ and $\eta$. For $D=0$ the I-N phase transition takes place at $T_{IN} = T_n^* + b_n^2/(4a_n c_n)$ and the N-SmA transition at $T_{NA} = T_s^* < T_{IN}$. In this case the phase transitions are determined by the nematic ($f_n^{(c)}$) and smectic ($f_s^{(c)}$) condensation term, respectively. The nematic elastic term $f_n^{(e)}$ enforces a spatially homogeneous ordering of $\mathbf{Q}(\mathbf{r})$, where $L$ stands for a positive representative nematic elastic constant. The term $f_n^{(f)}$ describes the coupling of an external electric field $\mathbf{E}$ with $\underline{Q}(\mathbf{r})$. For a positive field anisotropy $\Delta\varepsilon$ this term enforces parallel alignment of $\mathbf{n}$ and $\mathbf{E}$. The smectic elastic term $f_s^{(e)}$ is weighted by the positive smectic bend ($C_\perp$) and compressibility elastic constant ($C_\parallel$). The former tends to align a smectic layer normal along $\mathbf{n}$. Furthermore, the compressibility term enforces the smectic layer periodicity $q_0$. The coupling term $f_c$ could quantitatively and also qualitatively affect LC phase behavior.

**Minimal ODIC model**

We next consider the simplest possible mesoscopic modeling of plastic crystals. We refer to the corresponding minimal approach as the ODIC model. In the modeling, we originate from the minimal Landau-de Gennes model, which is designed to describe the sequence of phases, where first the nematic orientational order, and afterward additional translational order appears on decreasing temperature from the isotropic (liquid) phase. In the ODIC model, we assume the following sequence of 1st order phase transitions on decreasing temperature: liquid (L), plastic (P), crystal (C) phase. Long-range translational order is established at the L-P transition. In the subsequent P-C transition an additional orientational order also appears, which we describe by a vector order parameter. Therefore, the sequence in which ordering in orientational and translational ordering appears are in these cases reversed. We also propose an explanation why the P-C transition is often replaced by a glass-type transformation.

Following LC example, we describe a translational ordering with the complex order parameter

$$\psi = m_t e^{i\phi}. \tag{A4}$$

The amplitude $m_t$ measures the degree of translational ordering and the phase $\phi$ determines the translational structural ordering. For example, a structure exhibiting a simple periodic undulation determined by the wave vector $\boldsymbol{q} = q\boldsymbol{e}_t$ (where $|\boldsymbol{e}_t| = 1$), is described by

$$\phi = \boldsymbol{q}\cdot\boldsymbol{r} = q\boldsymbol{e}_t\cdot\boldsymbol{r}. \tag{A5}$$

We describe the orientational ordering with the vector

$$\boldsymbol{m} = m_o \boldsymbol{e}_o. \tag{A6}$$

where the amplitude $m_o$ measures the degree of orientational ordering, and the unit vector $\boldsymbol{e}_o$ points along a locally selected direction.

In terms of these fields we describe the phase sequence liquid-plastic-crystal phase on reducing the temperature. These phases are characterized by $\{m_o = 0, m_t = 0\}$, $\{m_o = 

$0, m_t > 0\}$, and $\{m_o > 0, m_t > 0\}$, respectively. Using a standard Landau-type phenomenological approach we write the free energy density $f$ as an expansion in order parameters $\psi$ and $\boldsymbol{m}$, where symmetry allowed terms are considered. We express it as a sum $f = f_t + f_o + f_c$ containing only translational ($f_t$) and orientational ($f_o$) degrees of freedom, and the term ($f_c$) describing the coupling between these degrees. We further decompose $f_d = f_d^{(c)} + f_d^{(e)} + f_d^{(f)}$ (where the subscript $_d$ stands either for $_t$ or $_o$) into the sum of the condensation ($f_d^{(c)}$), elastic ($f_d^{(e)}$), external field ($f_d^{(f)}$) contribution. These terms are in the lowest order expansion necessary to describe sequence of 1st order phase transitions L-P and P-C expressed as follows:

$$f_t^{(c)} = a_t(T - T_t^*)|\psi|^2 - b_t|\psi|^4 + c_t|\psi|^6, \tag{A7a}$$

$$f_t^{(e)} = \kappa_t|(i\boldsymbol{q}_0 - \nabla)\psi|^2, \tag{A7b}$$

$$f_t^{(f)} = -\chi_t^{(1)}|\psi|^2|\boldsymbol{E}_{eff}|^2 - \chi_t^{(2)}|\nabla\psi \cdot \boldsymbol{E}_{eff}|^2, \tag{A7c}$$

$$f_o^{(c)} = a_o(T - T_o^*)|\boldsymbol{m}|^2 - b_o|\boldsymbol{m}|^4 + c_o|\boldsymbol{m}|^6, \tag{A7d}$$

$$f_o^{(e)} = \kappa_o|\nabla\boldsymbol{m}|^2, \tag{A7e}$$

$$f_o^{(f)} = -\chi_o \boldsymbol{m} \cdot \boldsymbol{E}_{eff}, \tag{A7f}$$

$$f_c = -d_1|\boldsymbol{m}|^2|\psi|^2 - d_2|\boldsymbol{m} \cdot \nabla\psi|^2. \tag{A7g}$$

The quantities $a_t, b_t, c_t, a_o, b_o, c_o$ are positive material constants enabling a 1st order L-P and P-C phase transitions. If the order parameters are decoupled and in absence of external fields (i.e., $f_o^{(f)} = f_t^{(f)} = f_c = 0$), then $T_t^*$ and $T_o^*$ determine the supercooling liquid and plastic phase temperature, respectively. We set $T_o^* < T_t^*$. The translational elastic term for a positive translational elastic module $\kappa_t$ enforces $\phi = \boldsymbol{q}_0 \cdot \boldsymbol{r}$ and uniform value of $m_t$, where the wave vector $q_0$ determines the translational symmetry in the plastic phase. Note the difference between the translational elastic terms describing plastic and SmA phase ordering (see Eq. (A3e) and Eq. (A7b)). In LCs the nematic orientational order introduces a symmetry breaking direction along which translational order is established in the SmA phase. On the contrary in the plastic phase, the orientational order is absent. The quantity

$$\boldsymbol{E}_{eff} = \boldsymbol{E} + \boldsymbol{E}_{rf} = E_{eff}\boldsymbol{e}_{eff} \tag{A8}$$

stands for an effective electric field, which we decompose into an ordering ($\boldsymbol{E} = E\boldsymbol{e}_E$) and random field-type ($\boldsymbol{E}_{rf} = E_{rf}\boldsymbol{e}_{rf}$) component, and $\{\boldsymbol{e}_{eff}, \boldsymbol{e}_E, \boldsymbol{e}_{rf}\}$ are unit vectors. Here $\boldsymbol{E}$ determines a field contribution due to an applied external voltage to a cell confining a sample. If the constants $\chi_t^{(1)}$ and $\chi_t^{(2)}$ are positive, $\vec{E}$ promotes the formation of translational ordering. Due to symmetry consideration these terms are in the lowest order proportional with $E_{eff}^2$. The term weighted with $\chi_t^{(1)}$ ($\chi_t^{(2)}$) enforces isotropic (anisotropic) translational ordering if $\boldsymbol{E} \neq 0$. In cases, where a crystal phase is replaced by a short-ranged glass-like phase, we assume a strong enough value of $\boldsymbol{E}_{rf}$. The latter contribution could arise due to spatially nonuniform orientation of molecular electric dipoles within a system. Note that in general both the amplitude $E_{eff}$ and $\boldsymbol{e}_{eff}$ might exhibit random variations.

A positive orientational elastic module $\kappa_o$ favors homogeneous (i.e. $m_o$ is spatially uniform) and uniform orientational ordering along a symmetry breaking direction. The vector character of the orientational order parameter allows linear coupling with external field. The constant $\chi_o$ determines the coupling strength with an effective electric field $\boldsymbol{E}_{eff}$. For a positive value, it tends to locally align $\boldsymbol{m}$ along the field direction.

The term $f_c$ couples translational and orientational degrees of freedom. Due to symmetry requirements, it is in the lowest order expansion in order parameters proportional to $m_o^2 m_t^2$. We assume that both coupling constants $d_1$ and $d_2$ are positive, promoting mutual appearance of both degrees of ordering. The term proportional to $d_1$ ($d_2$) is isotropic (anisotropic).

We next consider a case when a constant external electric field $E$ is applied and focus on the $E$-driven pretransitional response. We set that the coupling of $E$ with the orientational degrees is stronger than with the translational order. Consequently, we neglect the free energy contribution given by Eq. (A7c). We also assume that $E \gg E_{rf}$. We set the spatially homogeneous order parameter. Furthermore, we assume that the dielectric response is dominated by orientational order, consequently we focus on $m_o(T)$ behavior.

For $E>0$ the most important free energy density $f$ contributions for temperatures $T > T_o^*$ are

$$f \sim (a_o(T - T_o^*) - m_t^2(d_1 + d_2 q_0^2))m_o^2 - \chi_o E m_o . \tag{A9}$$

Minimization of $f$ with respect to $m_o$ yields

$$m_o = \frac{\chi_o E}{2a_o\left(T - T_o^{(eff)}\right)} , \tag{A10}$$

where $T_o^{(eff)} = T_o^* + m_t^2(d_1 + d_2 q_0^2)/a_0$. For relatively weakly coupled order parameters it holds $T_o^{(eff)} = T_o^*$ for $T > T_t^{(c)}$, where $T_t^{(c)} = T_t^* + b_t^2/(4a_t c_t)$ determines the plastic-liquid phase transition temperature for $E=0$.

On the other hand for $T > T_o^{(c)}$, where $T_o^{(c)} = T_o^* + b_o^2/(4a_o c_o)$ determines the plastic crystal phase transition temperature for $E=0$, it roughly holds $T_o^{(eff)} \sim T_o^* + \frac{b_t(d_1 + d_2 q_0^2)}{2 c_t a_o}$.

Therefore, $T_o^{(eff)}$ has different values in the temperature regimes $T_t^{(c)} > T > T_o^{(c)}$ and $T > T_o^{(c)}$. Next we discuss conditions for which glass-type structures are expected. In the analysis above we neglected presence of random fields. These are in our treatment mimicked by $\boldsymbol{E}_{rf} = E_{rf} \boldsymbol{e}_{rf}$, see Eq.(A8). Their presence might have strong effects on the P-C phase transition. Namely, this transition corresponds to the continuous symmetry breaking in orientational ordering, which is directly coupled with $\boldsymbol{E}_{rf}$ (see Eq. (A7f)). According to the Imry-Ma theorem,[61,62] one of the pivotal theorems of statistical mechanics of disordered systems, even an infinitesimally weak random field-type disorder breaks long-range order due to the presence of Goldstone fluctuations in *m*. These are inevitable present due to the continuous symmetry breaking. The resulting configurations are expected to exhibit short-range order and well-characterized by a single domain length. However, recent studies[62] suggest that finite field strength is needed to establish short-range order in systems of our interest.

### APPENDIX B: CLAUSSIUS-MOSSOTI EQUATION

Here we recall the derivation of the classical Claussius-Mossoti equation which relates macroscopic dielectric response, represented by $\varepsilon$, with microscopic dipolar properties, represented by the dipolar polarisability $\alpha$.[45,59,60] We consider spatially homogeneous dielectric system, whose total electric dipole polarization **P** equals zero in the absence of an external electric field **E**. For the sake of simplicity, we represent the dielectric response with the scalar (i.e., orientationally independent) dielectric constant $\varepsilon$. In presence of **E** the macroscopic polarisation appears:

$$\boldsymbol{P} = \varepsilon_0(\varepsilon - 1)\boldsymbol{E}. \tag{B1}$$

Furthermore, it holds $\sim n\boldsymbol{p}$, where $\boldsymbol{p}$ stands for an average electric dipole and *n* is the volume density of dipoles. For relatively weak external dielectric fields one assumes a linear-type response

$$\boldsymbol{p} \sim \alpha \boldsymbol{E}_{loc}, \tag{B2}$$

where $\alpha$ is the polarizability constant, and

$$\boldsymbol{E}_{loc} \sim \boldsymbol{E} + \frac{\boldsymbol{P}}{3\varepsilon_0} \tag{B3}$$

is the local field acting on $\boldsymbol{p}$. By combining Eq. (B1), Eq. (B2) and Eq. (B3) one obtains the classical Claussius-Mossotti equation

$$\frac{\varepsilon-1}{\varepsilon+2} = \frac{N\alpha}{3\varepsilon_0}, \tag{B4}$$

which relates ε and α.

The Lorentz-Lorentz local field (Eq. (B3)) leads to the relation:

$$\varepsilon - 1 = \frac{N\alpha/\varepsilon_0}{1 - N\alpha/3\varepsilon_0} \tag{B5}$$

The Lorentz-Lorentz local field And then the Clausius-Mosotti equation are for non-polar dielectrics, for which the catastrophe emerging from Eq. B5 has to be absent. However, the catastrophe can appear for dipolar dielectrics where $\alpha = \alpha_e + \alpha_a + N\mu^2/3k_BT$, what yield the relation:

$$\varepsilon - 1 = \frac{3T_C}{T - T_C} \tag{B6}$$

with $T_C = \left(N\mu^2/9\varepsilon_0 k_B\right)\left[1/\left(1 - N(\alpha_e + \alpha_a)/3\varepsilon_0\right)\right]$

The Mosotti Catastrophe suggest the possible Curie-Weiss pretransitional behavior for classical non-ferroelectric dielectrics. For instance for water Eq. (6) leads to the para- to ferroelectric phase transition and solidification at $T_C \approx 1520K$, in clear disagreement with the nature. To overcome the paradox more realistic local fields were developed.[45,59,60]

However, the question arises what happens if a system of non-interacting or weakly interacting permanent dipole moment appears to be experimentally accessible? Is this the case of ODIC-forming materials?

**ACKNOWLEDGEMENTS**

This research was carried out due to the support of the National Centre for Science (Poland), project NCN OPUS ref. 2016/21/B/ST3/02203, head Aleksandra Drozd-Rzoska.